\documentclass[pre,twocolumn, superscriptaddress, amsmath, showpacs, nofootinbib, floatfix] {revtex4}

\usepackage[mathcal]{eucal}
\usepackage{bm}
\usepackage{graphicx}
\newcommand{\myvs}{\vspace{0.3cm}}
\setkeys{Gin}{width=1\linewidth}

\begin{document}

\title{Probability distribution of magnetization in the
one-dimensional Ising model:\\ Effects of boundary conditions}

\author{T. Antal} \email{antalt@cs.elte.hu}
\affiliation{ Physics Department,
  Simon Fraser University, Burnaby, BC V5A 1S6, Canada}
\affiliation{Institute for Theoretical Physics, E\"otv\"os University,
  1117 Budapest, P\'azm\'any s\'et\'any 1/a, Hungary}

\author{M. Droz}
\email{Michel.Droz@physics.unige.ch}
 \affiliation{D\'epartement de Physique Th\'eorique,
  Universit\'e de Gen\`eve, CH 1211 Gen\`eve 4, Switzerland}

\author{Z. R\'acz}
\email{racz@poe.elte.hu} \affiliation{Institute for
  Theoretical Physics, E\"otv\"os University, 1117 Budapest,
  P\'azm\'any s\'et\'any 1/a, Hungary}


\pacs{05.50.+q, 05.70.Jk, 24.60.-k, 75.10.Hk}

\begin{abstract}
Finite-size scaling functions are investigated both for the
mean-square magnetization fluctuations and for the probability
distribution of the magnetization in the one-dimensional Ising
model. The scaling functions are evaluated in the limit of the
temperature going to zero ($T\rightarrow 0$), the size of the system
going to infinity ($N\to\infty$) while $N[1-\tanh(J/k_BT)]$ is
kept finite ($J$ being the nearest neighbor coupling).  Exact
calculations using various boundary conditions (periodic,
antiperiodic, free, block) demonstrate explicitly how the scaling functions
depend on the boundary conditions. We also show that the block 
(small part of a large system) magnetization
distribution results are identical to those obtained for
free boundary conditions.
\end{abstract}

\maketitle

\section{Introduction}
Finite-size scaling has been developed intensively during the last
few decades \cite{{Fisher-fss},{Barber},{Cardy-fss},{Privman}} and it has
become a standard tool in the studies of critical systems. An
interesting application of the method is using the finite-size scaling
of the distribution function of the order-parameter fluctuations as
hallmarks of universality classes. The idea goes back to Bruce
\cite{{Bruce81},{Bruce85},{Bruce88}} who used it, {\it e.g.}, to
verify that the gas-liquid transition of the two-dimensional Lennard-Jones
fluid belongs to the Ising universality class
\cite{Bruce-Wilding}. The list of applications to thermodynamic
critical points is long \cite{equi-dist} and the idea has reemerged in
nonequilibrium critical (or effectively critical) systems
\cite{{nonequi-dist},{turbu-dist},{other-dist},{1f}}, as well.

The usefulness of scaling functions as hallmarks of universality
classes depends on their availability for comparisons.  Indeed, a
significant portion of the applications is about the Ising
universality class where the scaling functions for the $d=2,3,4$
dimensional distribution functions are well known from simulations
\cite{{Binder81},{Tsyp-Blote}} and field-theoretic results in
$d=4-\epsilon$ are also available \cite{Eisenriegler}. The picture
gallery of scaling functions is clearly far from complete since these
functions have been systematically worked out only for surface growth
models \cite{nonequi-dist} and for Gaussian $1/f^\alpha$ type noise
processes \cite{1f}.

An important issue concerning the critical distribution functions is
their dependence on the boundary conditions (BC). While theoretical
calculations usually address periodic systems, the building of a
histogram of a physical quantity ({\it e.g.}, the magnetization) in an
experimental system involves measuring the magnetizations in
patches of a given size within the bulk of the system (corresponding to
block-spin magnetizations in an Ising model).  As has been shown by
Binder \cite{Binder81}, the block-spin distribution function at
criticality depends on the BC on the block, a finding that is not
entirely unexpected since the infinite-range critical correlations
feel the boundaries of the system.

The BC dependence of the scaling functions is an interesting problem
and, indeed, there has been a series of works where the PDF of the
magnetization for the $d=2$ Ising model at the critical point has been
investigated for various BC \cite{Wu} including some exotic ones
(M\"obius strip, Klein bottle). Similar problems have also been
studied for the roughness distribution of $1/f^\alpha$ type noise
processes \cite{1f}.  Analytical results about BC dependence of the
scaling functions are scarce: they are restricted to Gaussian models
\cite{1f}, expansions around $d=4$ \cite{Eisenriegler} and around the
spherical limit \cite{Brezin}. This is why we decided to revisit the
the $d=1$ Ising model where the effect of BC can be seen in
analytical detail.

Although the critical temperature of the $d=1$ Ising model is zero, it
displays nontrivial features in its finite-size scaling as the
critical point is approached. The particular case of the distribution
function of the magnetization in bulk blocks has already been discussed by Bruce
\cite{Bruce81}. The purpose of this paper is to calculate the scaling
function for the case of periodic (PBC), antiperiodic (APBC), free (FBC)
and block (BBC) boundary conditions, and thus gauge the importance of
the role played by the BC.

For pedagogical purposes,
we also compute the finite-size scaling of the magnetization
fluctuations. The calculation is elementary in
this case and one can easily observe that the periodic, antiperiodic, and
free BC yield distinct scaling functions. Furthermore, one
can also see explicitly how the scaling function associated with the
block BC emerges when a small part of a
large system is used for measuring the fluctuations.

The evaluation
of the magnetization distribution is somewhat more involved but
can be carried out relatively simply by using a combinatorial approach.
For the periodic and free BC, the calculations
yield nontrivial functions which
are combinations of two delta peaks and a continuum background,
with the relative weight of the delta functions reduced for the case of FBC.
The delta functions disappear entirely for antiperiodic BC. Finally,
the combinatorial approach reproduces Bruce's result for the BBC and it
turns out that the block scaling function is identical to that of the FBC case.

\section{Model and notation}
We consider the one-dimensional Ising chain of $N$ spins ($\sigma_i =
\pm 1$, $i = 1,..,N$) with ferromagnetic ($J>0$) coupling and with
various BC. The interaction energy of a given
configuration of spins $\{\sigma_i\}$  is given by
\begin{equation}
  E(\{\sigma_i\}) = -J\sum_{i=1}^{N-1} \sigma_i \sigma_{i+1}
  - J_{bc} \sigma_N \sigma_1 ~,
\end{equation}
where $J_{bc}=J,\, -J,\, 0$ for periodic, antiperiodic and free BC
respectively.  The model is exactly solvable using, {\it e.g.}, the
transfer matrix formalism \cite{kw}, and the partition functions for
periodic BC (upper signs) and for antiperiodic BC (lower signs) are
\begin{equation}
  Z^{(p),(a)} = 2^N ( \cosh^N{K} \pm \sinh^N{K}) \, .
  \label{Zp}
\end{equation}
while the correlations are given by
\begin{equation}
  \langle \sigma_i\sigma_{i+n}\rangle^{(p),(a)}=
  \frac{v^n\pm v^{N-n}}{1\pm v^N} ~,
  \label{pcorr}
\end{equation}
where $v = \tanh{K}$ with $K=J/k_BT$ and, furthermore,
$1\le i,\, i+n \le N$.
The above quantities are particularly simple for free BC
\begin{equation}
  Z^{(f)} = 2^N\cosh^{N-1}{K}~, \qquad
  \langle \sigma_i\sigma_{i+n}\rangle^{(f)} =v^n ~.
  \label{fcorr}
\end{equation}
Note that the correlations only depend on the distance of the two
spins and not on their particular positions within the chain. This is
true not only in the periodic case but also for APBC and FBC.  Also
note that the $(p), (a)$, and $(f)$ superscripts refer to periodic,
antiperiodic and free BC respectively throughout the paper.

\section{Finite-size scaling of fluctuations}

The mean square fluctuations
of the total magnetization $M=\sum_{i=1}^N \sigma_i$
can be calculated via the spin correlations as
\begin{equation}
  \langle M^2\rangle
  =\sum_{i,j=1}^N\langle \sigma_i\sigma_{j}\rangle \, .
  \label{m2gen}
\end{equation}
We begin with the discussion of $\langle M^2\rangle$ for the periodic-,
free-, and antiperiodic BC, leaving the case of block BC for a separate
subsection.

\subsection{Periodic-, free-, and antiperiodic BC}

Substituting the expressions for the spin correlations
(\ref{pcorr},\ref{fcorr}) into Eq.(\ref{m2gen}), one easily finds
\begin{equation}
\frac{\langle M^2\rangle}{N}=
\begin{cases} 
 \displaystyle
 \frac{1+v}{1-v}\cdot \frac{1-v^N}{1+v^N} & \mbox{ PBC } \myvs\\
 \displaystyle
 \frac{1+v}{1-v}\cdot \frac{1+v^N}{1-v^N}
   -\frac{4v}{N(1-v)^2} & \mbox{ APBC} \myvs\\
 \displaystyle
 \frac{1+v}{1-v}-\frac{2v(1-v^N)}{N(1-v)^2} &\mbox{ FBC } ~.
 \end{cases}
\label{M2}
\end{equation}
In the thermodynamic limit ($N\to \infty$) all the above expressions
reduce to
\begin{equation}
\frac{\langle M^2\rangle}{N}=
\frac{1+v}{1-v} ~.
\label{M22}
\end{equation}
One can see that the fluctuations diverge as $T\to0$, and they have a
power-law singularity provided
$t=1-v=1-\tanh{J/k_BT}$ is used as the control parameter.
One can also read off (\ref{M22})
the value ($\gamma=1$) of the susceptibility exponent \cite{Baxter}.

The correlation-length exponent, $\nu$, is another exponent needed
in finite-size scaling. It is obtained from the
spin-spin correlations which decay exponentially in the thermodynamic limit.
The correlation length, $\xi$, defined by the exponential decay is independent
of the BC
\begin{equation}
 \langle \sigma_i\sigma_{i+n}\rangle_{N\to \infty}
 = v^n=e^{-n\ln{(1/v)}} =e^{-{n/\xi}} \, .
 \label{corrfunc}
\end{equation}
and diverges for $T\to 0$ as
\begin{equation}
 \xi =-\frac{1}{\ln{[1-(1-v)]}}\sim {(1-v)^{-1}}=t^{-1}
  \label{kszi}
\end{equation}
thus providing us with $\nu=1$.
As noted by Chen and Domb \cite{chen}, the above definition of the
correlation length (rather than that found from the second-moment of the
correlation function) is
the appropriate one to discuss finite size scaling for this model.

The scaling form one expects in finite-size scaling
in dimension $d$ is as follows \cite{Fisher-fss}
\begin{equation}
  \frac{\langle M^2\rangle}{N^{1+\gamma/d\nu}}=
\Phi(t N^{1/d\nu})
  \label{FSS}
\end{equation}
with $N$ being the number of spins.
Putting in the above expression $d=\gamma=\nu=1$,
the finite-size scaling suggestion takes the form
\begin{equation}
  \frac{\langle M^2\rangle}{N^{2}}=\langle m^2\rangle =
\Phi(t N)
  \label{FSSd1}
\end{equation}
where $m=M/N$ is the magnetization density.

\begin{figure}[htb]
\includegraphics[width=6cm,angle=-90]{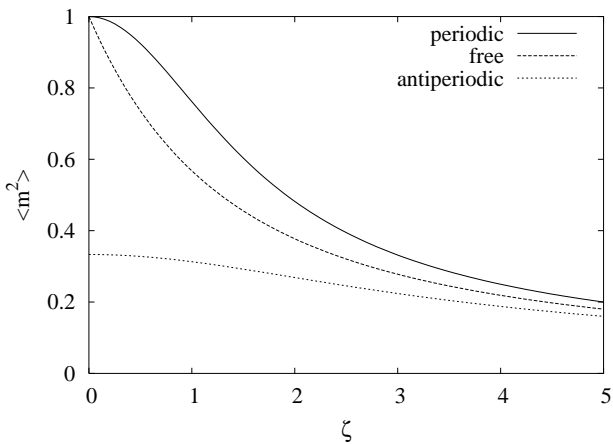}
\caption{Magnetization fluctuations in the scaling limit for various
  BC [see Eq.~(\ref{M2scaling})]. The
  scaling variable is $\zeta=Nt/2$.}
\label{fig:pfm2}
\end{figure}

Dividing both sides of Eqs.~(\ref{M2}) by $N$, we indeed find
that $\langle M^2\rangle/ N^2=\langle m^2\rangle$
yields well defined scaling functions
in the limit of  $t=1-v\to 0$ while
$t N=2\zeta$ is kept finite. The
scaling variable $\zeta$ has a simple meaning since, in the scaling limit
\begin{equation}
2\zeta=  Nt = N/\xi ~,
  \label{zetaxi0}
\end{equation}
{\it i.e.}, $2\zeta$ is the average number of domains (or domain walls)
in the system. The actual scaling functions are given below
\begin{equation}
\frac{\langle M^2\rangle}{N^2}=\langle m^2\rangle=
\begin{cases}
\displaystyle
~ \frac{1}{\zeta}\tanh{\zeta}&\mbox{ PBC } \myvs\\
\displaystyle
~ \frac{1}{\zeta}\left(\coth{\zeta}-\frac{1}{\zeta}\right)
&\mbox{ APBC} \myvs\\
\displaystyle
~ \frac{1}{\zeta}\left(1-\frac{1-e^{-2\zeta}}{2\zeta} \right)
&\mbox{ FBC,}
\end{cases}
\label{M2scaling}
\end{equation}
and they are shown on Fig.~\ref{fig:pfm2}. The functions belonging to
different BC are clearly distinct thus demonstrating
explicitly the BC dependence of the scaling functions.

The scaling functions coincide for $\zeta\to \infty$. This is understandable
since large $\zeta$ means large number of domain walls which
means that the disordered regime is approached where the effects of the
BC diminish. The differences in $\zeta\to 0$ limit can be
accounted for by the number of domain walls in the system. In particular,
the ground state is completely aligned for PBC and FBC systems while it
contains an arbitrarily positioned single domain wall in case of APBC.
As a result
\begin{equation}
\langle m^2\rangle^{(p),(f)}=1 ~,
\qquad \langle m^2\rangle^{(a)}=1/3
\label{m2zeta=0}
\end{equation}
explaining the values of the scaling functions at $\zeta= 0$.

The small $\zeta= 0$ behaviour of $\langle m^2\rangle$ can be
understood in terms of
the smallest energy excitations above the ground states.
These excitations are obtained from the ground state by
adding a pair
domain walls in case of PBC and APBC, while they consist of
a single domain wall for FBC. Their effect is a
quadratic (linear) decrease of $\langle m^2\rangle$ near $\zeta= 0$ for
PBC and APBC (FBC) systems.

\begin{figure}[htb]
\includegraphics[width=6cm,angle=-90]{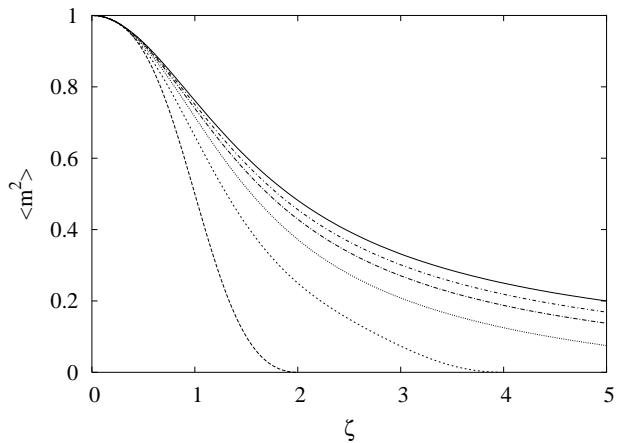}
\caption{Correction to scaling for magnetization fluctuations for
  periodic BC. First equation in (\ref{M2}) is used to calculate
  $\langle m^2\rangle=\langle M^2\rangle/N^2$ for system with $N=2, 4,
  8, 16, 32$, and $\infty$. The scaling variable is $\zeta=Nt/2$.The
  $1/N$ dependence of the correction can be visually observed (note
  that the distance from the $N=\infty$ curve is halved as $N$ is
  doubled)}
\label{fig:d1m2sc}
\end{figure}

We close this section with a note on the speed of convergence of the
scaling function. Looking at Eq.(\ref{M2}), one can see that $ \langle
M^2\rangle/N$ converges exponentially to its thermodynamic limit for
PBC, while the convergence is only power law $(N^{-1})$ for FBC and
APBC. As to the scaling function, it can be easily shown that the
convergence is power law $(N^{-1})$ even in PBC case. More precisely,
the correction term is of the form $N^{-1}g(\zeta)$ where $g(\zeta)\le
0$ with $g(\zeta\to 0)=0$ and $g(\zeta\to \infty)=-1$.
Fig.~\ref{fig:d1m2sc} displays $\langle m^2 \rangle$ calculated from
Eq.~(\ref{M2}) for the PBC case (note that for finite $N$ and using
the scaling variable $\zeta$, the results are meaningful only for
$\zeta<N$).  It should be clear from Fig.~\ref{fig:d1m2sc} that the
convergence is slow although it may appear to be quite good at small
$\zeta$ due to the particular form of $g(\zeta)$.

\subsection{Block (window) boundary conditions}

When studying magnetization fluctuations in an experiment,
one usually divides the system into blocks
and measures the block magnetizations. The corresponding
theoretical construction in the $d=1$ Ising model is to
consider a block (window) of length $\ell$ at a position $k$ in a system
of total length $N$,
and study the fluctuations of the block magnetization defined as
\begin{equation}
  \langle M_\ell^2\rangle
  =\sum_{i,j=k}^{k+\ell-1}\langle \sigma_i\sigma_{j}\rangle ~.
\label{M2block}
\end{equation}
Provided the full length of the block is within the system
({\it i.e.}, it does not contain the boundary with the coupling $J_{bc}$),
the correlations entering Eq.(\ref{M2block}) and, consequently,
$\langle M^2_\ell\rangle$ does not depend on the location of the block,
$k$, and we can write
\begin{equation}
  \langle M_\ell^2\rangle =
  \ell + 2 \sum\limits_{n=1}^{\ell-1}(\ell-n)
  \langle \sigma_1 \sigma_{1+n} \rangle  ~.
\end{equation}
Substituting now the correlations for PBC (\ref{pcorr}),
one readily obtains an expression which depends both on the
block size $\ell$ and the system size $N$:
\begin{equation}
 \frac{ \langle M_\ell^2\rangle^{(p)}}{\ell^2}=
 \frac{1-v^N}{1+v^N}\cdot\frac{1+v}{\ell(1-v)}
 -\frac{2v(1-v^{N-\ell})}{1+v^N}\cdot\frac{1-v^\ell}{\ell^2(1-v)^2}
\label{M2window2}
\end{equation}
where the superscript $(p)$ denotes the periodic BC.
Introducing the "aspect ratio" $b=\ell/N$, and using the scaling
variable $\zeta=\ell(1-v)/2$, the scaling limit $N\to \infty$, $\ell \to \infty$
with $b$ and $\zeta$ finite yields the following scaling function
\begin{eqnarray}
  \frac{\langle  M_\ell^2\rangle^{(p)}}{\ell^2} && =
  \langle m_\ell^2 \rangle^{(p)}(\zeta,b)=\\
  &&\frac{1}{\zeta}\left( \tanh{\zeta/b} -
  \frac{1-e^{-2\zeta}}{2\zeta}
  \cdot \frac{1-e^{-2(1-b)\zeta/b}}{1+e^{-2\zeta/b}}\right) \nonumber\, .
  \label{Windowsc}
\end{eqnarray}
As one can see, the scaling function goes over into the FBC
case if $b\to 0$ while it becomes the scaling function for the
PBC case if $b=1$. The function smoothly interpolates between the
limiting cases in $0< b\le1$.

If the block is embedded in a FBC system then
$\langle M_\ell^2 \rangle^{(f)}$ can be deduced
from the observation that $\langle \sigma_1 \sigma_{1+n} \rangle$
is independent of the system size (see Eq.~\ref{fcorr}). Namely,
$\langle M_\ell^2 \rangle^{(f)}$ must coincide with
$\langle M^2 \rangle$ for the FBC with $N$ replaced by $\ell$.
Thus we have a scaling function (\ref{M2scaling}) which
does not depend on the "aspect ratio" $b$
\begin{equation}
  \frac{ \langle M_\ell^2\rangle^{(f)}}{\ell^2}=
  \langle m_\ell^2\rangle^{(f)}(\zeta,b)=
  \langle m^2\rangle^{(f)}(\zeta) \, .
  \label{Window sc-fbc}
\end{equation}

Finally, the fluctuations in a block embedded in an antiperiodic chain,
can also be calculated and the somewhat more complicated result
has a similar structure as in case of embedding in a periodic chain. Namely,
changing the aspect ratio from $b=1$ to $b=0$, the result interpolates
between the APBC and the FBC scaling functions in Eq.(\ref{M2}).

The common feature of all the above results is that the BC 
become irrelevant in
the limit $b\to0$ where the block is much smaller than the system.
Furthermore, we find that this bulk behaviour coincides with the FBC
result. Whether this coincidence with the FBC result is a general feature
of bulk fluctuations is not quite clear and should be
investigated in more complicated and higher dimensional systems.

\section{Magnetization distribution}

If at a given temperature, {\it i.e.}, at a given correlation length, the
system size goes to infinity then the magnetization distribution goes
to a Gaussian around zero due to the Central Limit Theorem, and goes
eventually to a Dirac delta function
\begin{equation}
\lim_{T\to0} \lim_{N\to\infty} P(m) = \delta(0) ~.
\end{equation}
On the other hand, at any given system size as the temperature goes to
zero, {\it i.e.}, the correlation length goes to infinity, the magnetization
goes to either plus or minus one, and the distribution becomes
\begin{equation}
\lim_{N\to\infty} \lim_{T\to0} P(m) =  \frac{1}{2}[\delta(m+1)+\delta(m-1)] ~.
\end{equation}
The importance of the $\zeta=Nt/2$ scaling variable is that if
$N\to\infty$ and $T\to0$ in such a way that the
correlation length is always proportional to the system size ({\it
i.e.}, $\zeta$ is a constant (\ref{zetaxi0})) then a nontrivial
distribution arises even for the $d=1$ Ising model.

The meaning of $\zeta$ suggests the
development of a small $\zeta$
(small number $g$ of domain walls) expansion.
Thus we shall first calculate
$P_g(M)$, the probability of a given magnetization
in the presence of $g$ domain walls.
Once $P_g(M)$ is known, the
probability of a given magnetization $P(M)$ can be obtained
by summing up $P_g(M)$
\begin{equation}
  P(M) = \sum_g P_g(M) ~.
  \label{Psum}
\end{equation}
The states with $g$ domain
walls are degenerate as their energy $E_g$ depends only on the number of
walls
\begin{equation}
 E_g =
 \begin{cases}
 J(2g-N) &\mbox{ PBC, APBC} \myvs\\ 
 J(2g-N+1) &\mbox{ FBC } ~.
 \end{cases}
\label{Eg}
\end{equation}
Thus, the calculation $P_g(M)$
reduces to counting all possible spin configurations
$\Omega_g(M)$ at given $g$ and $M$ values
\begin{equation}
  P_g(M) = \frac{e^{-E_g/k_BT}}{Z} \Omega_g(M) ~.
\label{PMbase}
\end{equation}
Here $Z$ is the partition function corresponding to a given BC
which, in the scaling limit, becomes
\begin{equation}
Z =
\begin{cases}
 { 2 e^{NK}\cosh{\zeta}}&\mbox{ PBC } \myvs\\
 { 2 e^{NK}\sinh{\zeta}} &\mbox{ APBC} \myvs\\
 2  e^{(N-1)K} e^\zeta &\mbox{ FBC.}
\end{cases}
\label{Zlimit}
\end{equation}
For $g=0$, {\it i.e.}, if there is no domain wall in the
configurations
\begin{equation}
  \Omega_0(M) = \delta_{M, N} + \delta_{M, -N} ~.  
  \label{C0Mdef} 
\end{equation}

For $g>0$, let $W_j$ be the position of the $j$th wall (say $W_j=i$
when it is just before $\sigma_i$) with $j=1,..,g$, {\it i.e.},
$W_j<W_{j+1}$ for $j=1,..,g-1$, and also note that by definition
$W_1>0$ and $W_g\le N$.  Let us denote the magnetization of the
domain between $W_1$ and $W_2$ by $M_1$. Now, it is sufficient to obtain
the number of configurations with the restriction $M_1>0$ and denote
it as $\Omega_g^+(M)$ as
\begin{equation}
  \Omega_g(M) = \Omega_g^+(M) + \Omega_g^+(-M) ~.  
  \label{CsMcomb}
\end{equation}
Note that fixed magnetization also means fixed number of upspins 
$N_\uparrow=(N+M)/2$ and downspins $N_\downarrow=(N-M)/2$.

\subsection{Periodic boundary conditions}

PBC enforce the number of domain walls to be always even $g=2s$, with
$s=0,...,[N/2]$, where $[~]$ stands for the integer part. As there is
at least one spin after each domain wall, we can imagine those spins
as being attached to the walls. A typical configuration looks like
\begin{equation}
  \downarrow\downarrow\downarrow(W_1\uparrow)\uparrow\uparrow
  (W_2\downarrow)\downarrow\downarrow\downarrow \dots
  \uparrow\uparrow
  (W_{2s}\downarrow)\downarrow\downarrow ~.
  \label{demo}
\end{equation}
Now counting all the possible configurations is equivalent to
distributing in all possible ways the $N_\uparrow-s$ (not attached)
upspins among the $s$ up domains (those domains where the spins are up,
{\it i.e.}, between $W_{2j-1}$ and $W_{2j}$) and independently distributing
the $N_\downarrow-s$ (not attached) downspins among the $s+1$ down
domains (between $W_{2j}$ and $W_{2j+1}$, also in front of $W_1$ and behind
$W_{2s}$)
\begin{equation}
  \Omega_{2s}^+(M) = \binom{N_\uparrow-1}{s-1} \binom{N_\downarrow}{s} ~.
  \label{omega+}
\end{equation}
Note that the binomial coefficient $\binom{a}{b}=0$ for $a<b$, which
reflects the fact that there are no configurations with more domain
walls than either $2N_\uparrow$ or $2N_\downarrow$, {\it i.e.},
$s\le\min(N_\uparrow,N_\downarrow)$. With the formula (\ref{CsMcomb}) one
can easily arrive at $\Omega_{2s}$ without restriction on the sign
of $M_1$
\begin{equation}
  \Omega_{2s}(M) = \binom{N_\uparrow-1}{s-1} \binom{N_\downarrow}{s}
  + \binom{N_\downarrow-1}{s-1} \binom{N_\uparrow}{s} ~.
  \label{omegadisc}
\end{equation}

In the $N\to\infty$ limit, for
fix $n_\uparrow=N_\uparrow/N$ and $n_\downarrow=N_\downarrow/N$ the
number of configurations becomes
\begin{equation}
  \Omega_{2s}(M) = N^{2s-1} \frac{(n_\uparrow n_\downarrow)^{s-1}}
  {s!(s-1)!} ~.
  \label{omegaper}
\end{equation}
In this limit we also need to switch from the discrete probabilities
$P(M)$ to the probability density $P(m)$, which brings in a factor
$N/2$, and substituting Eq.~(\ref{omegaper}) into Eq.~(\ref{PMbase})
leads to
\begin{equation}
  P_{2s}(m) = \frac{e^{NK}}{2Z^{(p)}} e^{-4sK} N^{2s} 
  (n_\uparrow n_\downarrow)^{s-1} ~.
  \label{Psminter}
\end{equation}
Now the scaling limit can be finally taken using (\ref{Zlimit}), and
for $s=0$ one realizes that the expression of Eq.~(\ref{C0Mdef})
develops singularities
\begin{equation}
  P_0(m) = \frac{1}{2\cosh\zeta} 
  \left[ \delta(m+1) +  \delta(m-1) \right] ~.
  \label{P0mresult}
\end{equation}
For $s\ne 0$, using $\zeta = N e^{-2K}$ leads to the final
result for the magnetization distribution with a fixed $2s$ number of
walls
\begin{equation}
  P_{2s}(m) = 
  \frac{(\zeta/2)^{2s}(1-m^2)^{s-1}}{s!(s-1)!\cosh{\zeta}}  ~.
  \label{Psmresult}
\end{equation}
It becomes clear at this point that we are doing a small $\zeta$ expansion
and that a fixed $2s$ number of domain walls belongs to the order $2s$ 
of the expansion.

\begin{figure}[htb]
\includegraphics[width=6cm,angle=-90]{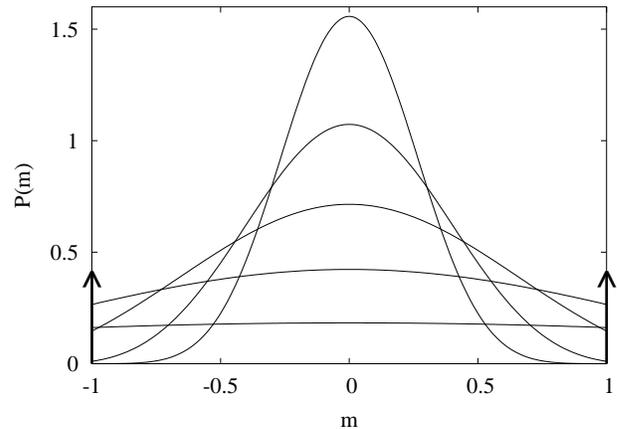}
\caption{Magnetization distribution $P(m)$ in the $d=1$ Ising model
  with periodic BC in the $T\to 0$ and $N\to\infty$
  limit as a function of $m=M/N$ with the scaling variable
  $\zeta=Nt/2$ fixed at values $\zeta=1$, 2, 4, 8,
  16 (from bottom at origin). The arrows symbolize the singular parts
  $\sim\delta(m\pm1)$ of the distributions of Eq.~(\ref{Pmresult}). }
\label{fig:pdis}
\end{figure}

The magnetization distribution without restrictions on the number of
walls can be obtained from Eq.~(\ref{Psum}) which is valid also in the
continuous limit
\begin{equation}
  P(m) = \sum_{s=0}^\infty P_{2s}(m) ~.
  \label{Pmsump}
\end{equation}
Using the series form of the modified Bessel functions
\begin{equation}
  I_\nu(x) = \sum_{k=0}^\infty \frac{(x/2)^{2k+\nu}}{k!(k+\nu)!}
  \label{bessel}
\end{equation}
leads to the final expression for periodic BC
\begin{eqnarray}
  \label{Pmresult}
  P^{(p)}(m) &=&  \frac{1}{2\cosh\zeta} 
  \left[ \delta(m+1) +  \delta(m-1) \right]\\ 
  &+& \frac{\zeta}{2\sqrt{1-m^2}\cosh{\zeta}} 
  I_1\left(\zeta\sqrt{1-m^2}\right) ~. \nonumber
\end{eqnarray}

One can easily check that $P(m)$ is normalized, {\it i.e.},
$\int_{-1}^1 dm~P(m)=1$, by changing the integration variable 
$m$ to $\theta=\arcsin m$ and using the series form of Eq.~(\ref{bessel}).
In the same way the expression for the second moment of $P(m)$
in Eq.~(\ref{M2scaling}) can also be obtained.

\begin{figure}[htb]
\includegraphics[width=6cm,angle=-90]{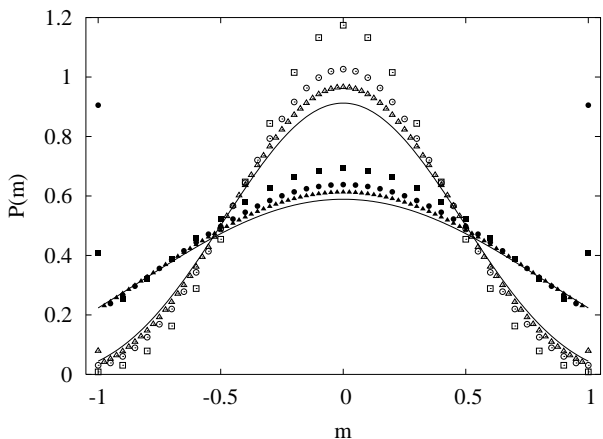}
\caption{Magnetization distribution $P(m)$ for periodic BC in the
  $N\to\infty$ limit [only the regular part of Eq.~(\ref{Pmresult}) is
  depicted] compared to its finite $N=20$ (square), 40 (circle), and
  80 (triangle) forms of Eq.~(\ref{Pmfinite}) with the scaling
  variable being $\zeta= 3$ (closed symbols), and 6 (open symbols).
  Observe the evolving singularities at $m=\pm1$ for finite systems
  (closed triangles are out of range).}
\label{fig:d1PM}
\end{figure}

One can investigate the speed of convergence in $N$ of the
magnetization distribution $P(m)$ of Eq.~(\ref{Pmresult}). Instead of
making Monte Carlo simulations on the Ising model we calculated numerically
the probabilities of possible magnetizations for finite chains, based
on Eq.~(\ref{PMbase}) and (\ref{Pmsump}), and multiplied the results
by $(N+1)/2$ for the sake of comparison with the probability density
$P(m)$ in the scaling limit
\begin{equation}
  P(m,N) = \frac{N+1}{2Z}\sum_{s=0}^{[N/2]} e^{(N-4s)K} \Omega_{2s}(M) ~,
  \label{Pmfinite}
\end{equation}
where $\Omega_{2s}(M)$ is given by Eq.~(\ref{C0Mdef}),
(\ref{omegadisc}), and $K=\mbox{artanh}(1-2\zeta/N)$. Eq.~(\ref{Pmfinite})
can be easily evaluated numerically, as the the computation time
increases linearly with $N$ as opposed to the exponentially long time
needed to encounter all possible configurations.  One observes in
Fig.~\ref{fig:d1PM} that the convergence is faster for smaller values
of $\zeta$, in agreement with Fig.~\ref{fig:d1m2sc}.

\subsection{Antiperiodic boundary conditions}

The main difference from the periodic case is that the number of
domain walls is odd $g=2s-1$, with $s=1,...,[(N+1)/2]$. As there has
to be at least one domain wall in each configuration there are no
Dirac delta peaks at $m=\pm1$ in the distribution. A typical
configuration looks like
\begin{equation}
  \downarrow\downarrow(W_1\uparrow)\uparrow\uparrow\uparrow
  (W_2\downarrow)\downarrow\downarrow \dots
  \downarrow\downarrow
  (W_{2s-1}\uparrow)\uparrow\uparrow\uparrow ~.
  \label{demoanti}
\end{equation}
The number of configurations with the restriction of $M_1>0$ is
\begin{equation}
  \Omega_{2s-1}^+(M) = \binom{N_\uparrow-1}{s-1} \binom{N_\downarrow}{s-1}~,
  \label{omega+anti}
\end{equation}
as we have $N_\uparrow-s$ free upspins to distribute into $s$ up
domains and $N_\downarrow-s+1$ free downspins for $s$ down domains.
In the continuous limit the number of configurations becomes
\begin{equation}
  \Omega_{2s-1}(M) = 2N^{2s-2} \frac{(n_\uparrow n_\downarrow)^{s-1}}
  {(s-1)!^2} ~,
  \label{omegaanti}
\end{equation}
and the magnetization distribution with fixed
number of walls (\ref{PMbase}) reads
\begin{equation}
  P_{2s-1}(m) = 
  \frac{(\zeta/2)^{2s-1}(1-m^2)^{s-1}}{(s-1)!^2\sinh{\zeta}}  ~.
  \label{Psmresultanti}
\end{equation}
Summing this expression up over the possible number of walls (\ref{Psum})
leads to the final result
\begin{equation}
  \label{Pmresultanti}
  P^{(a)}(m) =  
  \frac{\zeta}{2\sinh \zeta}
  I_0\left(\zeta\sqrt{1-m^2}\right) ~.
\end{equation}

\subsection{Free boundary conditions}

In case of free BC the number of domain walls
can be both even $g=2s$ and odd $g=2s-1$ with $s=1,...,[N/2]$.
The only difference in counting all possible configurations, with a
given $g$, $M$, and the condition $M_1>0$, is the supplementary
restriction that there always has to be a down spin before the first
wall for obvious reasons, which can be visualized as $(\downarrow
W_1\uparrow)$ in the example (\ref{demo}).  
Now the number of configurations can be easily obtained
\begin{eqnarray}
  \Omega_{2s-1}^+(M) &=& \binom{N_\uparrow-1}{s-1}
  \binom{N_\downarrow-1}{s-1} \\  
  \Omega_{2s}^+(M) &=&
  \binom{N_\uparrow-1}{s-1} \binom{N_\downarrow-1}{s} \nonumber ~,
\end{eqnarray}
and one observes that in the scaling limit they are equal to the
corresponding periodic (\ref{omegaper}) or antiperiodic
(\ref{omegaanti}) result, {\it i.e.},
$\Omega_{2s}^{(f)}(M)=\Omega_{2s}^{(p)}(M)$ and
$\Omega_{2s-1}^{(f)}(M)=\Omega_{2s-1}^{(a)}(M)$.  The energies of both
the even and odd states are greater than the energies of the
corresponding periodic and antiperiodic states (\ref{Eg}) by $J$ and
thus we can write the scaling limit of the probability of a given $M$
as
\begin{eqnarray}
  \displaystyle
  P_g^{(f)}(M) &=&\frac{e^{-E_g^{(f)}/k_BT}}{Z^{(f)}} \Omega_g^{(f)}(M) \myvs\\
   &=&
  \begin{cases}\nonumber
  \displaystyle
  \frac{e^{-K}Z^{(p)}}{Z^{(f)}}P_g^{(p)}(M)& \qquad g=2s \myvs\\
  \displaystyle
  \frac{e^{-K}Z^{(a)}}{Z^{(f)}}P_{g}^{(a)}(M)& \qquad g=2s-1 .
  \end{cases}
\label{PMfree}
\end{eqnarray}
Using Eqs.(\ref{Zlimit}), one finds that the prefactors of the
distributions depend only on $\zeta$, thus collecting the
contributions with different number of domain walls (\ref{Psum}), we
obtain an expression for $P^{(f)}(m)$ through $P^{(p),(a)}(m)$
\begin{equation}
 P^{(f)}(m) = e^{-\zeta} [ \cosh\zeta~ P^{(p)}(m) + \sinh\zeta~ P^{(a)}(m)]~,
\label{relation}
\end{equation}
which leads to
\begin{eqnarray}
  \label{Pmresultfree}
  P^{(f)}(m) &=&  \frac{e^{-\zeta}}{2}
  \left[ \delta(m+1) +  \delta(m-1) \right] \\
  &+& \frac{\zeta e^{-\zeta}}{2\sqrt{1-m^2}}
  I_1\left(\zeta\sqrt{1-m^2}\right) \nonumber \\
  &+& \frac{\zeta e^{-\zeta}}{2}
  I_0\left(\zeta\sqrt{1-m^2}\right) ~. \nonumber
\end{eqnarray}
As we shall see in the next section, the above expression applies for
bulk BC as well. The bulk result was obtained previously by Bruce (see
\cite{Bruce81} equation (3.20)). Note also that the relationship
established between the distributions for different BC
(\ref{relation}) is certainly valid for the fluctuations of the
magnetization (\ref{M2scaling}) as well.

\begin{figure}[htb]
\includegraphics[width=6cm,angle=-90]{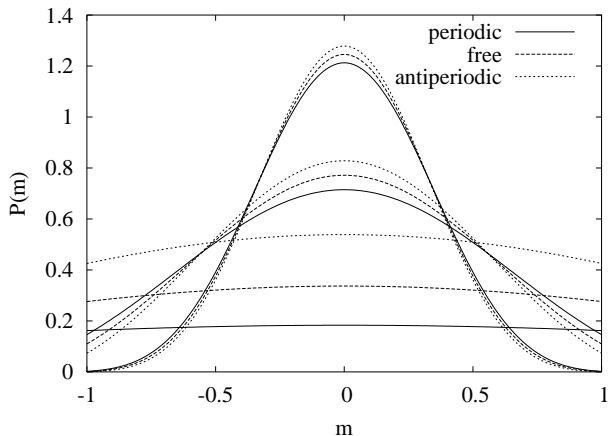}
\caption{Comparison of the regular parts of the magnetization
  distribution $P(m)$ for periodic, antiperiodic, and free BC
  in the scaling limit, with the scaling variable being
  $\zeta= 1$, 4, and 10 (from bottom at origin). Note that the Dirac
  deltas at $m=\pm1$ of the periodic and the free case are not
  displayed.}
\label{fig:dfdis}
\end{figure}

One should note that the coefficient of the singular part is larger
for the periodic case, {\it i.e.}, a periodic system is more likely to
be in the completely ordered steady state. This is expected due to the
lower energy of the states with $2s-1$ domain walls than that with
$2s$ domain walls. On Fig.\ref{fig:dfdis} we display the nonsingular
part of the distributions. This nonsingular part also shows the
expected sequence from PBC resulting in the most ordered state to APBC
yielding the most disordered state.

\subsection{Block boundary conditions}

We start with the simplest case, {\it i.e.}, impose free BC
on the chain and investigate the magnetization
of a finite segment (block) of length $\ell$.
As we shall show the distribution of $M_\ell$
is identical to the free end result for any $\ell$.  In order to see this,
consider the Boltzmann weight of a configuration
of the spins for the segment of spins between
$\sigma_k$ and $\sigma_{n=k+\ell}$
\begin{equation}
  P(\{\sigma_k,...,\sigma_n\})=
    \frac{\sum\limits_{\{\sigma\}_{1,k-1},
    \{\sigma\}_{n+1,N}}
    e^{K\sum_{j=1}^{N-1}\sigma_j\sigma_{j+1}}}
    {\sum\limits_{\quad\{ \sigma\}_{1,N}}
    e^{K\sum_{j=1}^{N-1}\sigma_j\sigma_{j+1}}}
  \label{def}
\end{equation}
where $\sum_{\{ \sigma\}_{i,j}}$ denotes summing over possible
values of the spins between sites $i$ and $j$.
One can "integrate out" the end spin $\sigma_1=\pm 1$ in both the numerator
and the denominator yielding cancelling factors $e^K+e^{-K}$.
This can be repeated till the spin $\sigma_k$ is reached and then
the same can be done starting from the other end ($\sigma_N$) of the chain.
As a result one obtains
\begin{equation}
  P(\{\sigma_k,...,\sigma_n\})= \frac {
  e^{K\sum_{j=k}^{n-1}\sigma_j\sigma_{j+1}} }
  {\sum\limits_{\quad\{\sigma\}_{k,n}}
  e^{K\sum_{j=k}^{n-1}\sigma_j\sigma_{j+1}}} ~.
\end{equation}
This is the free end probability distribution for the spins in the
segment $[k,k+\ell]$, thus the magnetization distribution $P(m_\ell)$,
with $m_\ell=M_\ell/\ell$, is identical to the FBC case given by
Eq.(\ref{Pmresultfree}), {\it i.e.},
$P^{(f)}(m_\ell) = P^{(f)}(m)$.

The above derivation does not hold for chains with PBC and APBC
and one expects that the $P(M_\ell)$ depends on the aspect ratio
$b=\ell/N$.  In the $b\to0$ limit,
{\it i.e.}, when the window size is relatively small, however,
$P(M_\ell)$ becomes independent from the BC imposed on
the whole chain.  More precisely, this happens in the scaling limit $\ell,
N\to\infty$, $t\to0$ with $\zeta=\ell t/2$ and $b=\ell/N$ kept
constant, a limit where the correlation length $\xi\sim\ell\ll N$ and the
effects of the boundaries of the chain can be neglected.

\subsection{Asymptotic regimes}

The small $\zeta$ expansion is already given by Eq~(\ref{Psum}). For
sufficiently small $\zeta$ the distribution $P(m)$ can be approximated
by the first few terms, {\it e.\ g.}, the Dirac delta functions (except for
the antiperiodic case) plus a constant probability density for all
BC.

For large values of $\zeta$ a Gaussian approximation can be obtained
using only the first term of the large $x$ asymptotic of the Bessel
functions
\begin{equation}
I_1(x) \approx I_0(x) \approx \frac{e^x}{\sqrt{2\pi x}},
\end{equation}
and using also the condition $m\ll 1$, where $P(m)$ significantly
differs from zero, due to the factor $\exp(\zeta\sqrt{1-m^2})$
\begin{equation}
P(m) \approx \frac{1}{\sqrt{2\pi\zeta}}e^{-m^2\zeta/2} ~.
\end{equation}
P(m) eventually evolves to a Dirac delta function $\delta(m)$ for
$\zeta\to\infty$.

\section{Final remarks}

The fluctuations of the magnetization in the one-dimensional Ising
model have been investigated near its critical point in the limit of
$T\to0$ and $N\to\infty$ with the average number of domain walls
($2\zeta=Nt$) kept constant.  A simple combinatorial derivation has
been presented for the magnetization distributions and it was showed
that in this limit these distributions are nontrivial well defined functions 
of the magnetization and the control parameter $\zeta$.

The $\zeta$ dependence of the magnetization distribution should be
emphasized, as it means that the order parameter distribution of a
system can, in general, depend on the way the critical point is
approached and the thermodynamic limit is taken.

We focused our attention on the effect of boundary conditions, namely
we imposed periodic, antiperiodic, and free BC on the chain.
The magnetization distributions are shown to be sensitive to the BC
and well distinguishable for all values of $\zeta$. 
For antiperiodic BC the distributions differ fundamentally from the 
periodic and free BC case (lack of Dirac delta peaks for APBC).

We also showed that the distribution of the magnetization of a segment
(BBC) of the whole chain with FBC coincide with the distribution of
the total magnetization, independently of the size of the segment.
For PBC or APBC chains the above statement is true only if the relative
segment size $b$ goes to zero.

It is worth mentioning the analogies to a simple random walk.  The
corresponding quantity is the distribution of the width of the walk
(mean square deviation of the position of the walker),
and for that two distinct functions have been exactly derived for periodic 
and free BC \cite{1f}. The qualitative behaviour of BBC is also
the same as that of the Ising model discussed above.

It would be very interesting to see similar calculations for the order
parameter distribution in other equilibrium one-dimensional models,
{\it e.g.}, the classical and quantum XY and Heisenberg models, or
nonequilibrium ones, {\it e.g.}, absorbing state phase transitions.

\section{Acknowledgment}
We would like to thank Z. Bajnok, G. Gy\"orgyi, and M. Plischke for useful
discussions.
This research has been partially supported by the Hungarian Academy
of Sciences (Grant No.\ OTKA T029792 and T043734) and by the Swiss
National Science Foundation.

\end{document}